# Integrating a QPSK Quantum Key Distribution Link

M. B. Costa e Silva[1], Q. Xu[1], S. Agnolini[1], S. Guilley[1], J-L. Danger[1], P. Gallion[1], F. J. Mendieta[2]

1 : Ecole Nationale Supérieure des Télécommunications (GET/Télécom Paris and CNRS),
46 rue Barrault, 75013 Paris, France. Email: mcosta@enst.fr
2 : on leave from CICESE, km.107 Carr. Tijuana, Ensenada, Baja California 22800, México

**Abstract** *We present the integration of the optical and electronic subsystems of a BB84-QKD fiber link. A high speed FPGA MODEM generates the random QPSK sequences for a fiber-optic delayed self-homodyne scheme using APD detectors.*

**Introduction**

In fiber-optic quantum key distribution (QKD) at telecommunications wavelengths, QPSK encoding of Alice's Qbits constitutes an attractive modulation format [1, 2] since it could be detected in differential interferometric schemes that need only one fiber, thus relaxing the requirements in phase and polarization stability and providing a more reliable long distance QKD link, and at the same time simplifies the system's integration and control.

When used with gated avalanche photon counters which operate in (Geiger) gate-mode at an optimal low temperature, the detected output must be properly processed, in order to compensate the relatively low quantum efficiency due to the quenching process.

In order to automate the QKD, a modem using Field Programmable Gate Array (FPGA) generates high speed random phases at a rate of 200Mbps.

**Experimental Work**

In our configuration Alice sends phase modulated key pulses at quantum levels, time-multiplexed with stronger unmodulated pulses in the same fiber, which constitutes a carrier phase reference. Alice encodes its symbols as antipodal phase states in two conjugated bases, resulting in a QPSK format, by using a two-electrode Mach-Zehnder electro-optical modulator (EOM-A)[3], allowing the independent choice of base and symbol as required by the BB84 protocol. When $\Phi_A = \pi/4$ or $-\pi/4$, the key is bit 0 in Base 1 or Base2, $\Phi_A = -3\pi/4$ or $3\pi/4$ represents the bit 1 in Base 1 or Base 2, respectively.

Figure 1 is a diagram of our experimental setup: optical pulses are fed into Alice's unbalanced interferometer: Alice's fainted key pulse QPSK signal is produced in the longer arm with an EOM-A, while unmodulated pulses pass through the shorter arm with accurate polarization control. Strong attenuation together with the losses in the longer arm (due to EOM-A as well as the connection between standard and maintain polarization fibers) produces modulated pulses at the average value of 0.1 photon/bit, therefore weaker than the unmodulated ones.

At the receiver, Bob's measurements are performed by applying his 2-state phase modulation ($\Phi_B = \pi/4$ or $-\pi/4$ as Base 1 or Base 2) in a similar delayed interferometric configuration so that the key pulses beat synchronously with the reference pulses, and are differentially photodetected using a pair of avalanche photon counters (APD) operating in the gated mode. Detector 1 clicks for $\Phi_A - \Phi_B = 0$ while Detector 2 clicks for $\Phi_A - \Phi_B = \pi$. When $\Phi_A - \Phi_B = \pi/2$ or $-\pi/2$, the photon arrives at Detector1 or Detector2 in a random way.

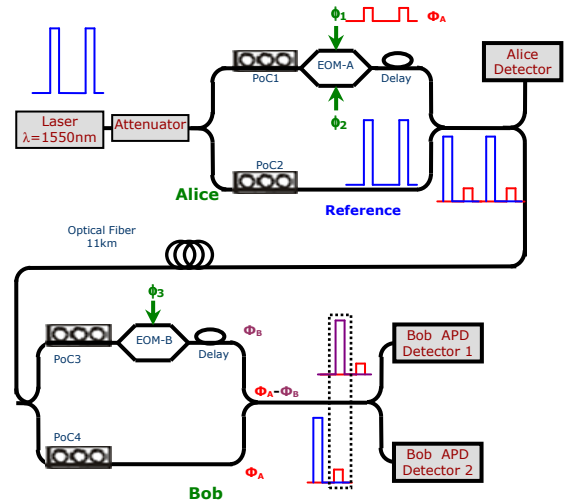

Fig.1 Delayed homodyne QKD photon counting setup

Bob received key pulses of approximately 0.1 photon/bit as well as stronger reference pulses when the laser source was strongly attenuated. The beating of the reference pulse and the key pulse increased the probability of detecting the key bit despite the presence of relatively noisier reference pulse which brought about some false counts. Since reference pulses were stronger than key pulses, more than 1 photon/bit was detectable at the APD detectors, at the repetition rate of 1MHz.

Figure 2 is the histogram of the measurement of a constant key bit where $\Phi_A - \Phi_B = 0$. The false clicks on

Detector 2 were due to the following causes: a) the unsatisfactory extinction ratio of the reference pulses; b) the polarization imperfections of the laser pulse which affected the mixing of the weaker key pulse with the strong reference pulse; c) the slight deviation of modulating signals; d) the unavoidable environmental variations; and e) the dark counts of the APD detectors. Using a fiber interferometer without feedback from the detected signal, the mean false count rate of 30% was obtained in a condition of stable phase for several minutes. A closed loop configuration would improve the performance.

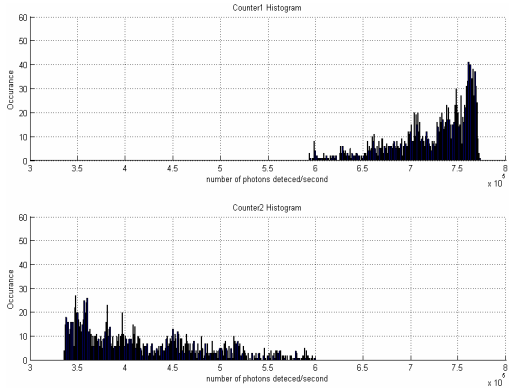

Fig. 2 Histogram of photon counting when $\Phi_A-\Phi_B = 0$

**Electronic subsystem**

In a view not to slow down the overall BB84 system, the random phases ($\Phi_1$, $\Phi_2$, and $\Phi_3$) must be provided at a rate possibly higher than the repetition rate. Moreover, it must be able to sustain burst buffer read from the network. The typical data rates are presented in Tab. 1.

Table 1 Data rates of the QKD system

| OSI Layer | Max. Speed | Limitation factor |
|---|---|---|
| Optics | 4Mbps | APD detectors |
| Electronics | 200Mbps | FPGA Max. Freq. |
| Network | 100Mbps | Ethernet 100-base T |

The synchronization of the system is implemented in both optical and electronic layers as shown in Fig 3. The phase drift is compensated by the optical signal feedback while the bit-level synchronization is delegated to the electronics, named here as MODEM. Consequently, a FPGA device is used to generate the random sequences at a constant throughput.

To achieve a random bit generation rate higher than 200Mbps, we exploit the D-Flip-Flop (DFF) metastability in open loop, in Fig. 4. A delay-line made up of small routing segments introducing a delay of $\Delta \approx 25$ ps, is built in the FPGA. Each segment is sampled by a DFF. The output word is then analyzed: the DFF that has the highest metastability serves as a seed for a Pseudo-RNG (Random Number Generator).

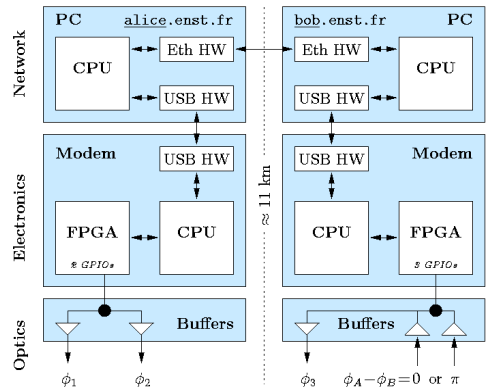

Fig. 3 General Architecture of the BB84 System

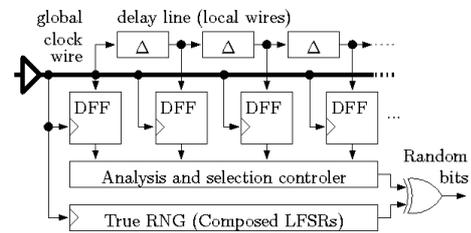

Fig. 4 Architecture of the random bit generation

The MODEM is connected to the network so that Alice and Bob can communicate their bases choices to extract the quantum keys. This transfer is asynchronous and possibly much faster than the quantum bit rate.

**Conclusions**

A telecom application of unconditional security has been designed. A properly designed high speed random phase generator (MODEM) controls a fiber-optic QKD system working in QPSK modulation scheme. APD photon counters are used with a differential scheme that reduces stability requirements, leading to a much easier and more stable long distance QKD system. Since the quantum efficiency of the APD detector is lower than 10%, the strong reference pulse offers a good method to improve the key detectability.